# Interactive Analysis of Static, Dynamic, and Crystalline SDTrimSP Simulations: Application to Nitrogen Ion Implantation into Vanadium

Miroslav Lebeda[a,b], Jan Drahokoupil[a], Vojtěch Smola[a], Petr Vlčák[a]

[a] *Faculty of Mechanical Engineering, Czech Technical University in Prague, Technická 4, 16607 Prague 6, Czech Republic*

[b] *Faculty of Nuclear Sciences and Physical Engineering, Czech Technical University in Prague, Trojanova 339/13, 12000 Prague 2, Czech Republic*

**Corresponding author:** Miroslav Lebeda, lebedmi2@cvut.cz

## Abstract

SDTrimSP is a widely used Monte Carlo simulation code based on the Binary Collision Approximation (BCA) for modeling ion implantation and ion-solid interaction processes. While an established graphical user interface (GUI) exists for simulation setup and execution, efficient post-processing, comparison of multiple simulations, and preparation of specific input file parameters remain limited. In this work, we present a web-based interface (sdtrimsp.app) that complements existing SDTrimSP tools by focusing on interactive visualization and analysis of depth distribution profiles. The platform enables direct upload and comparison of static and fluence-dependent dynamic profiles, supports unit conversion, and provides a calculator for determining the adjustable atomic density parameter of implanted ions required in dynamic simulations. In addition, the interface offers automated conversion of standard crystallographic file formats into the SDTrimSP-specific crystal structure input format (for version 7.00) for simulations into crystalline targets. The capabilities of the interface are demonstrated for nitrogen ion implantation into vanadium, including amorphous static and dynamic simulations and static crystalline simulations for different surface orientations. The results illustrate fluence-dependent saturation effects as well as orientation-dependent ion channeling behavior. Overall, the presented web-based tool provides a convenient and flexible extension to existing SDTrimSP workflows.

## 1. Introduction

SDTrimSP is a well-established Monte Carlo (MC) simulation code based on the Binary Collision Approximation (BCA), being widely used for modeling ion implantation and related ion-solid interaction processes[1]. Its applicability spans a broad range of research fields, including plasma-surface interactions, planetary science, semiconductor processing, surface modification, or radiation effects in materials[2–6]. Owing to its computational efficiency and flexibility, including the availability of multiple models for nuclear and electronic stopping, SDTrimSP is particularly well suited for investigating phenomena such as sputtering, the fluence-dependent evolution of depth distribution profiles in amorphous targets, or ion transport in crystalline materials, where crystallographic effects such as ion channeling can be characterized.

SDTrimSP simulations are defined through text-based input files, with the primary and necessary simulation parameters specified in the *tri.inp* file, being processed in a console environment as a Fortran code. To facilitate the use of SDTrimSP, a graphical user interface (GUI) has been recently developed[7], allowing to conveniently define simulation parameters, generate input files, directly execute simulations, and visualize selected output quantities. The GUI also provides plotting capabilities for SDTrimSP output files, which are well suited for immediate inspection of simulation results in a local desktop environment.

In many practical workflows, SDTrimSP simulations are executed on remote machines or high-performance computing systems (HPC)[8], while output files are transferred for analysis on different platforms or shared among collaborators. In such cases, post-processing and visualization directly within the existing GUI can be less convenient, particularly when comparing multiple simulations, fluence steps, or with external datasets such as from experiments or other simulation codes. Web-based tools that allow direct upload, visualization, and comparison of SDTrimSP output files without local installation would therefore offer a complementary and flexible approach.

In addition, fluence-dependent dynamic simulations require the determination of an adjustable atomic density parameter for the implanted element in order to correctly reproduce the density of the evolving target material. Similarly, simulations in crystalline targets require crystal structure input file in an SDTrimSP-specific format (*crystal.inp,* for latest main version of SDTrimSP 7.00). These tasks are not available in the current publicly available version of the SDTrimSP GUI (support for crystalline materials has only been added recently to the SDTrimSP itself) and are therefore typically handled outside the main simulation workflow using manual procedures. Providing dedicated and user-friendly support for these tasks can therefore notably streamline simulation setup and reduce the potential for user error.

In this work, we present a web-based interface that extends the existing SDTrimSP (available at sdtrimsp.app) tools by focusing on convenient post-processing, visualization, and specific tasks for the input file preparation. The interface enables interactive plotting and comparison of static and dynamic depth profiles upon uploading the SDTrimSP result into the application, calculation of the atomic density parameter of implanted ions for dynamic simulations, and conversion of commonly used crystallographic formats into the specific SDTrimSP crystal structure input format. The capabilities of the interface are demonstrated using low-energy (4 keV) nitrogen (N) ion implantation into vanadium (V), including amorphous static simulations, fluence-dependent dynamic simulations, and static crystalline simulations for the analysis of ion channeling effects. V and V-based materials are relevant for applications in nuclear technology and surface engineering, where N is commonly used to modify near-surface properties by forming vanadium nitrides (VN)[9–11]. The presented results therefore support the optimization of N implantation for controlled VN formation in V-based materials.

## 2. Implementation Details

The web-based interface is implemented in Python and deployed as an interactive application using the Streamlit framework (streamlit.io). It is accessible online via sdtrimsp.app and is currently hosted on the Streamlit Community Cloud, which enables direct deployment from a public GitHub repository (github.com/bracerino/sdtrimsp-output-plot). The repository documentation includes a tutorial for local installation. Running the application locally is recommended for users who wish to upload larger output files or perform more extensive analyses, as performance is then limited only by their local machine. Within the application, the Pymatgen library[12] is used to convert crystallographic information file (CIF) into POSCAR format, which is subsequently transformed into the SDTrimSP-specific crystal structure input file (*crystal.inp*). At present, the interactive interface is developed for SDtrimSP version 7.00 and has been tested with version 7.01.

## 3. SDTrimSP Simulation Setup

N ion implantation into the V target was simulated using SDTrimSP with static and dynamic mode into an amorphous target, and with static simulations into differently oriented surface of crystalline V phase. Input files were generated using the existing SDTrimSP GUI and subsequently modified to include also the calculations into the crystal.

The N ion energy was set to 4 keV. Surface binding energies and displacement energies were kept at their default values. Nuclear stopping was modeled using the Ziegler-Biersack-Littmark (ZBL) potential[13], while electronic stopping was described using the Ziegler-Biersack[14] model. Each simulation consisted of 1000 histories with 500 projectiles per history, and Gauss-Legendre integration was employed.

The atomic density of V was set to 0.072 atoms/Å$^3$, corresponding to the bcc phase, publicly available in the Crystallography Open Database (COD)[15] with ID 4001362[16]. For fluence-dependent dynamic simulations, the parameter representing the atomic density of N was calculated as follows: Assuming a limiting composition of 50 at. % N and 50 at.% V corresponding to the VN phase with an atomic density of 0.1209 atoms/Å$^3$ (available in the Materials Project (MP)[17] database with an ID 1002105), the N atomic density parameter was calculated as 0.3768 atoms/Å$^3$ following the SDTrimSP formalism. This calculation is implemented directly within the presented interface (see also **Chapter 4.1**). Furthermore, during the dynamic simulations, the maximum allowed fraction of implanted N was set to 51 at.%, permitting a slight exceedance of the ideal equiatomic VN composition.

For simulations of N into different surface orientations of crystalline V, the same physical parameters were used with a static mode. The crystal structure input file was generated using the conversion tool provided by the presented web-based application by uploading the aforementioned V structure file from the COD database.

## 4. Workflow and Capabilities of the Web-Based SDTrimSP Interface

### 4.1 Calculation of Atomic Density Parameter for Implanted Ions and Crystal Structure Conversion

In fluence-dependent dynamic SDTrimSP simulations, the atomic density of the implanted element behaves like an adjustable parameter which needs to be set in order to correctly reproduce the density of the evolving target material at a specific concentration of the implanted element and the target elements. For N implantation into V, a limiting composition of 50 at.% N and 50 at.% V was assumed. The corresponding target atomic density was therefore supposed as of the VN phase, 0.1209 atoms/Å$^3$. Based on this assumption, the required atomic density parameter for N was calculated as 0.3768 atoms/Å$^3$. The specific equation used for this calculation can be found in the official SDTrimSP documentation and is implemented directly within the presented web-based interface (**Figure 1**).

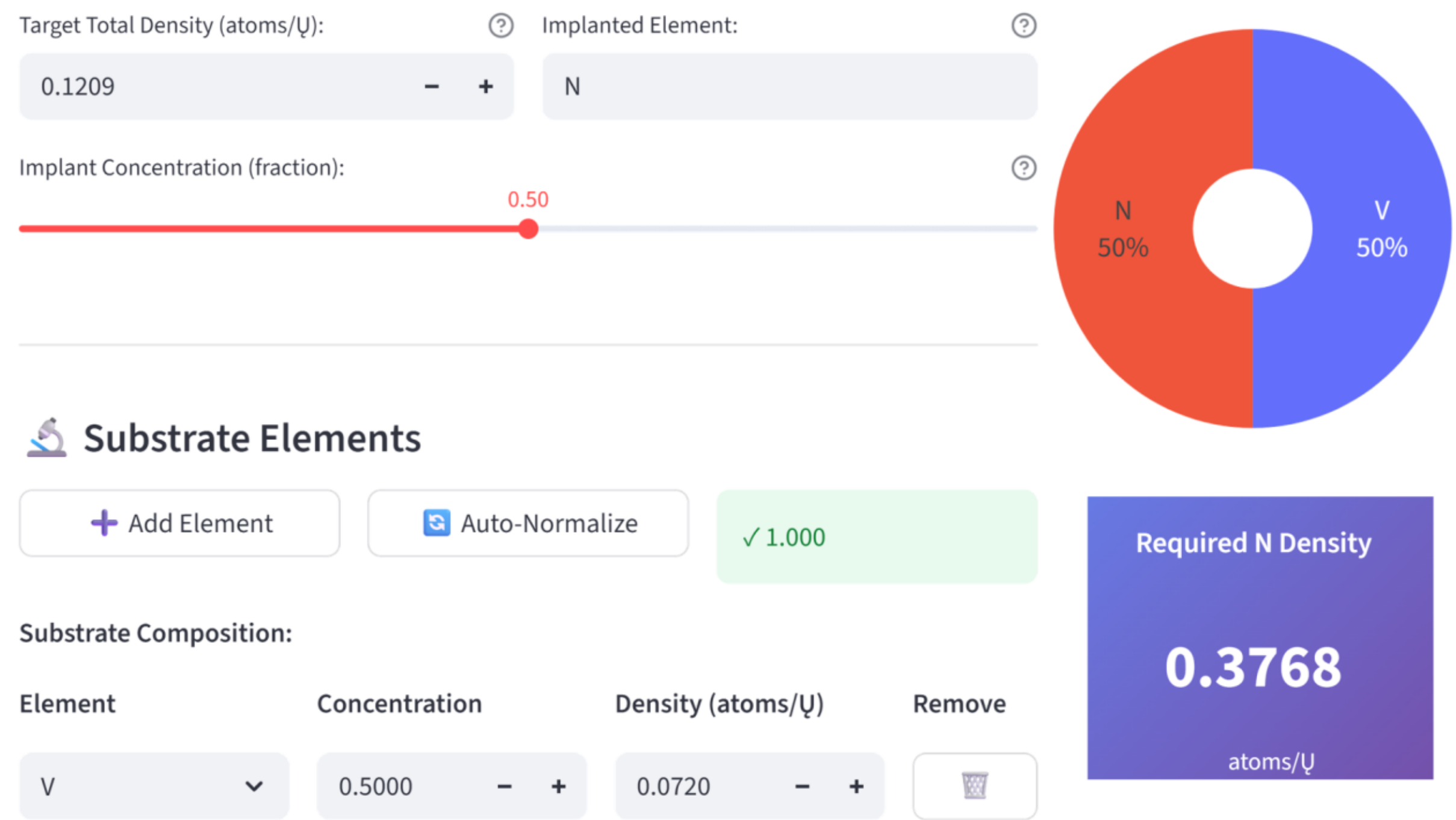

**Figure 1:** Calculation interface for determining the adjustable atomic density parameter of the implanted element in fluence-dependent dynamic SDTrimSP simulations. The example illustrates N implantation into V, where the N atomic density is calculated as 0.3768 atoms/Å$^3$ in order to reproduce the atomic density of the VN phase (0.1209 atoms/Å$^3$) at a limiting composition of 50 at.% N and 50 at.% V.

For simulations involving crystalline targets, SDTrimSP (version 7.00) requires a crystal structure input file (*crystal.inp*) describing primarily the lattice geometry, atomic positions and elements. In the present work, the required *crystal.inp* file for V was generated directly within the web-based interface by converting a standard CIF file of bcc V phase from the COD database with the ID 4001362. The interface currently supports POSCAR and CIF formats and performs the conversion automatically after the file is uploaded. This functionality streamlines the preparation of crystalline SDTrimSP simulations and reduces the need for manual preparation of a structure file.

Reorient crystal lattice vectors

Preview crystal.inp:

```
V
1
1
"V"
3.0287 0.0000 0.0000
0.0000 3.0287 0.0000
0.0000 0.0000 3.0287
2
0.0000 0.0000 0.0000 1
0.5000 0.5000 0.5000 1
0.0
0.0 0.0
5
```

Download crystal.inp

**Elements:** V

**Number of atoms:** 2

**Lattice vectors (Å):**

```
a = [3.0287, 0.0000, 0.0000]
b = [0.0000, 3.0287, 0.0000]
c = [0.0000, 0.0000, 3.0287]
```

**Figure 2:** Preview of the SDTrimSP crystal structure input file (*crystal.inp*) generated by the web-based interface from uploaded CIF file of bcc V phase.

## 4.2 Visualization and Comparison of Static and Dynamic Depth Profiles of Implanted N in V

Depth distribution profiles obtained from static simulations (*depth_proj.dat* or *depth_damage.proj*) and from fluence-dependent dynamic simulations (e.g., *E0_31_target.proj*) can be uploaded directly into the interface for interactive visualization and analysis. When a dynamic output file is loaded, the user can select a specific fluence step for display or visualize multiple fluence steps simultaneously to examine the evolution of the implantation profile. **Figure 3** illustrates the evolution of the N depth distributions (as directly provided by the web-based interface) in V for fluences of $1 \times 10^{16}$, $5 \times 10^{16}$, and $10 \times 10^{16}$ ions/cm$^2$. At the highest fluence, the depth profile exhibits a clear saturation effect resulting from the imposed maximum N concentration of 51 at.% in the V target. Moreover, the interface includes option for automatically determining concentration maxima across the depth profile and tracking their evolution as a function of fluence (**Figure 4**).

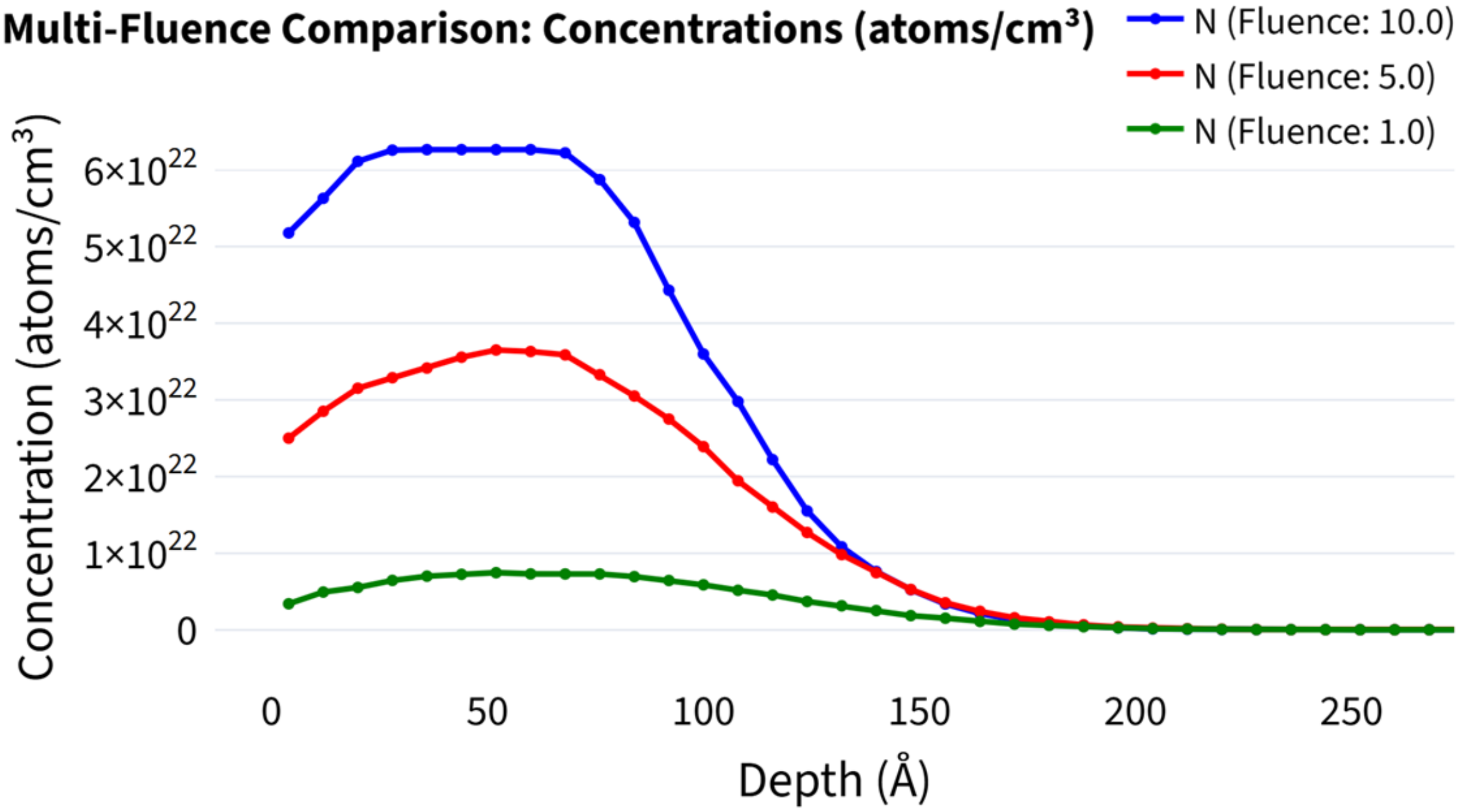

**Figure 3:** Evolution of N depth distributions in V obtained from fluence-dependent dynamic SDTrimSP simulations. Depth profiles are shown for fluences of $1 \times 10^{16}$, $5 \times 10^{16}$, and $10 \times 10^{16}$ ions/cm$^2$. At the highest fluence, the profile exhibits saturation effects resulting from the imposed maximum N concentration of 51 at.% in V.

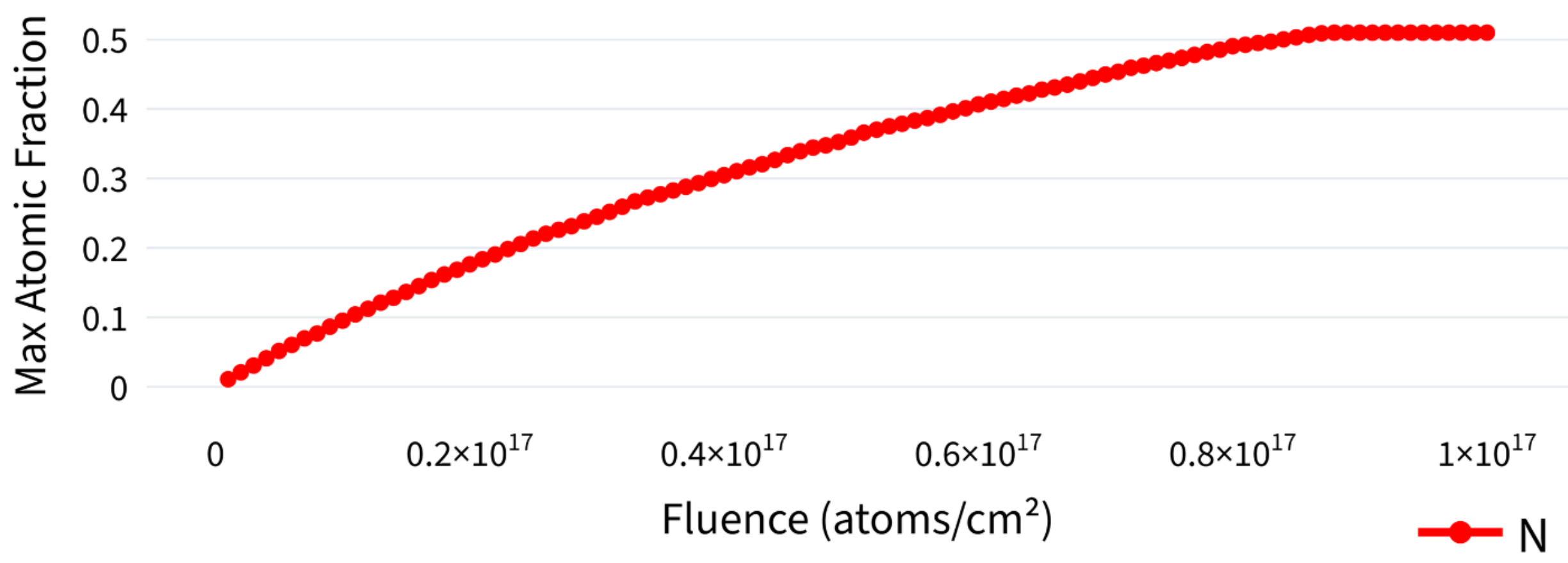

**Figure 4:** Evolution of the maximum N concentration in V as a function of implantation fluence, obtained from fluence-dependent dynamic SDTrimSP simulation.

For uploaded SDTrimSP output files, the interface offers options for unit conversion and data representation. The depth axis can be displayed in either ångströms or nanometers, while the concentration axis may be expressed as absolute atom counts, atomic probability, atomic fraction, or concentration in ions/$cm^3$. In addition, optional smoothing of depth profiles is supported using commonly applied techniques, including Savitzky-Golay filtering, moving-average smoothing, and Gaussian convolution.

The interface also supports the upload of external two-column datasets, such as experimentally measured depth profiles or results obtained from other simulation software. These datasets can be visualized alongside SDTrimSP results and directly compared within the same interactive plot. As an illustrative example, static amorphous simulation data were treated as external profiles and compared with dynamic simulation results at fluences of $1 \times 10^{16}$ and $10 \times 10^{16}$ ions/$cm^2$ (**Figure 5**). At lower fluence, the static and dynamic profiles show relatively good agreement, as expected. At higher fluence, however, dynamic simulations exhibit pronounced deviations due to changes in atomic density and concentration saturation, leading to profiles that shift toward the surface and accurately reflect the imposed maximum N concentration of 51 at.% in V.

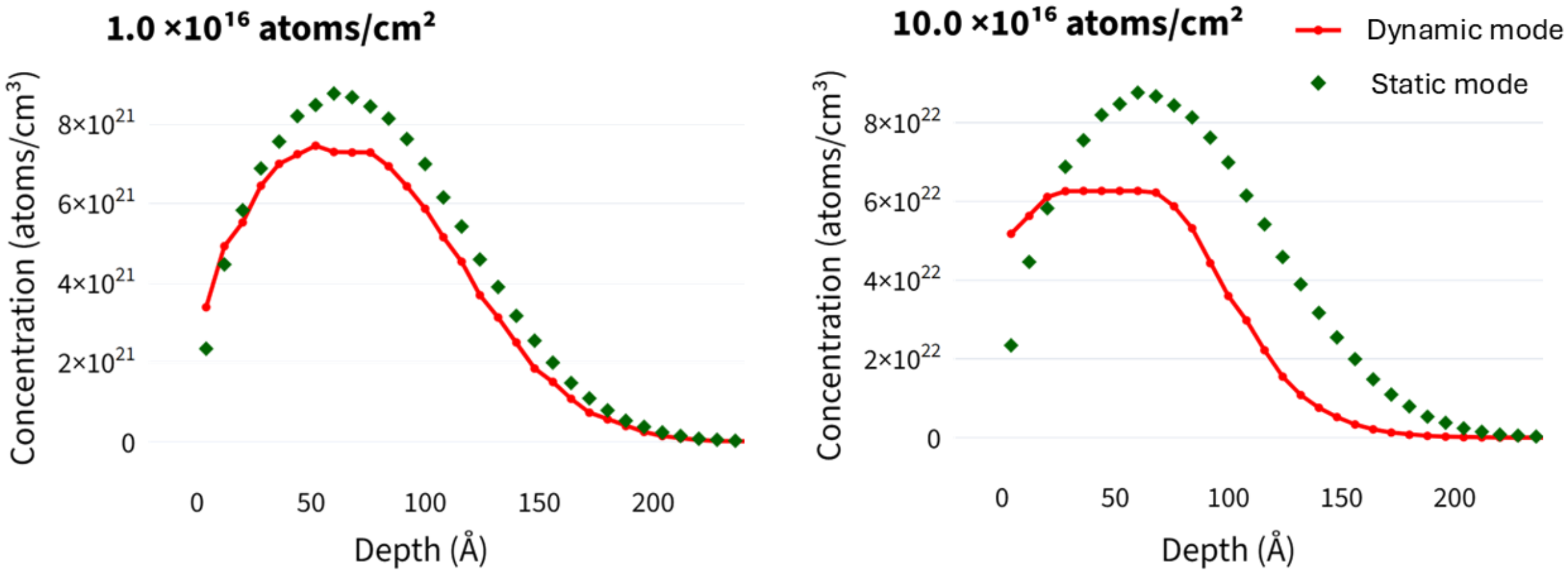

**Figure 5:** Comparison of N depth distributions in V between static and dynamic mode from SDTrimSP simulations. Fluence-dependent dynamic profiles are shown for fluences of $1 \times 10^{16}$ and $10 \times 10^{16}$ ions/$cm^2$ and are compared with the corresponding static profiles scaled for the same fluences.

## 4.3 N Ion Channeling in Crystalline V

For N implantation into crystalline V, depth distribution profiles were calculated for the following surface orientations: (100), (110), (111), (210), (211), (221), (321), and (12 7 5). **Figure 6** presents these orientations as views perpendicular to the surface normal, showcasing the potential presence of open crystallographic channels. These simulations were performed to characterize the influence of crystallographic orientation on the ion transport and to identify potential ion channeling effects in the resulting depth profiles.

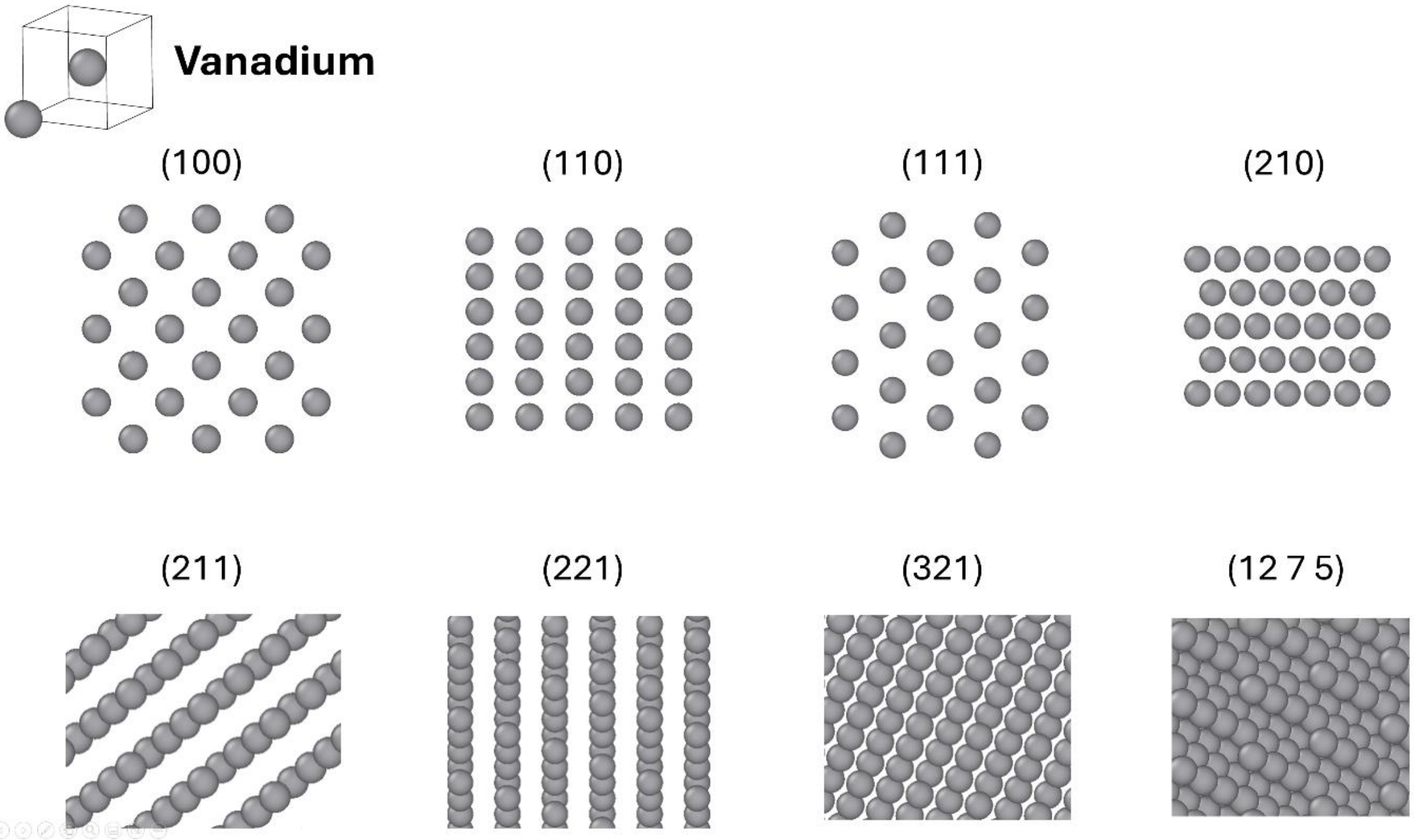

**Figure 6:** Surface orientations of the bcc V crystal used for N ion implantation simulations in the SDTrimSP crystal mode, shown as views perpendicular to the surface normal.

As shown in **Figure 7**, the most pronounced channeling effect is observed for the (111) orientation, where a sharp and well-defined channeling peak appears at significantly greater depth (close to 500 Å) compared with the non-channeling maxima located near the surface (close to 50 Å). This is in agreement with a general trend in bcc materials, where the (111) surface orientation was shown as the strongest channeling one[18,19]. A second pronounced channeling feature, characterized by a strong peak at increased penetration depth, is visible for the (100) orientation. The (211) and (221) orientations exhibit weaker channeling behavior, manifested as smaller secondary peaks at larger depths. Orientations (110) and (210) show enhanced penetration tails due to partial channeling. However, in these cases, the main profile maximum remains closer to the surface.

The high-index (12 7 5) orientation does not exhibit open channels and therefore shows no distinct channeling peak. Nevertheless, when compared with the amorphous static simulation, the depth profile still displays a small fraction of ions penetrating to greater depths, indicating residual orientation-dependent effects. Similarly, the depth distributions for the (321) orientation resemble the amorphous static profile more closely than those of low-index orientations, although noticeable differences remain, particularly in the penetration tail, where crystalline simulations show slightly enhanced ion transport relative to the amorphous case.

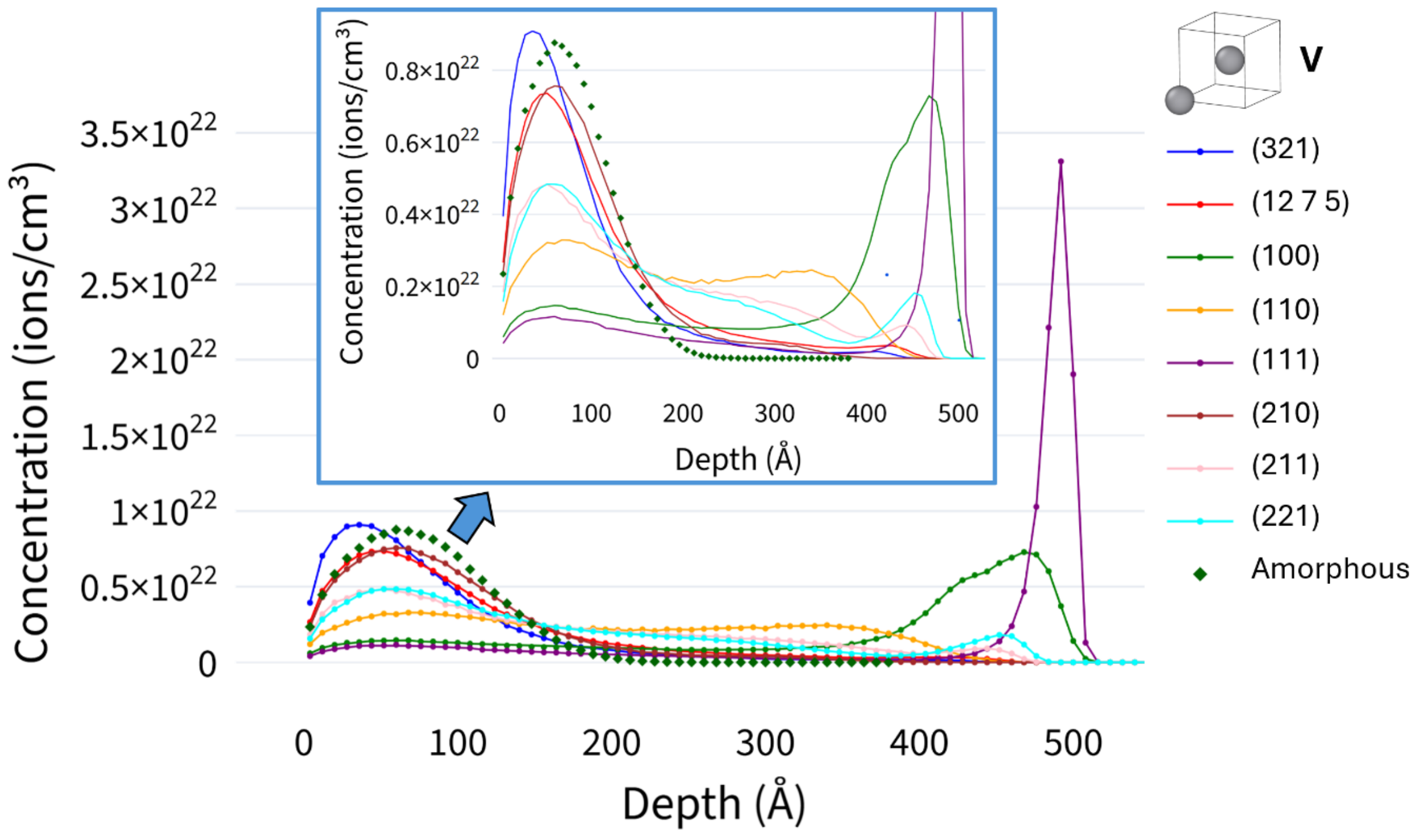

**Figure 7:** N depth distribution profiles in crystalline V obtained from static SDTrimSP simulations for different surface orientations, illustrating orientation-dependent ion channeling effects. For comparison, results from the amorphous target are also shown. The concentrations are scaled on the fluence of 1 × $10^{16}$ ions/cm$^2$.

## 5. Conclusions

In this work, we presented a web-based interface (sdtrimsp.app) that extends SDTrimSP and its native local GUI by providing an online platform for the convenient upload, visualization, and analysis of SDTrimSP output files. The interface enables interactive analysis of depth distribution profiles obtained from static, fluence-dependent dynamic, and crystalline implantation modes, supports direct comparison between static and dynamic simulations, and allows external datasets such as experimental profiles or results from other simulation codes to be uploaded for side-by-side comparison. In addition, the platform includes a tool for calculating the adjustable atomic density parameter of implanted ions required for fluence-dependent dynamic simulations. Automated conversion of standard crystallographic file formats into SDTrimSP-compatible crystal structure input files further simplifies the setup of simulations involving crystalline targets.

The applicability of the interface was demonstrated for N implantation into V, illustrating both fluence-dependent saturation effects in amorphous targets and pronounced orientation-dependent ion channeling phenomena in crystalline phase. By reducing manual effort and improving accessibility, the presented tool provides a practical and complementary extension to existing SDTrimSP workflows.

## 6. Acknowledgments

This work was supported by the Grant Agency of the Czech Technical University in Prague [grant No. SGS24/121/OHK2/3T/12].

## 7. Code Availability

The online version of the introduced interactive web-based application for SDTrimSP analysis is accessible at: sdtrimsp.app. The underlying code is publicly available on the GitHub repository for the local installation: github.com/bracerino/sdtrimsp-output-plot. Video tutorial how to use the application for the dynamic mode is available at: implant.fs.cvut.cz/sdtrimsp-depth-distribution-profiles. Another tutorial for calculating the ion implantation into the crystalline phase to observe ion channelling and plot its results in the interactive application is available at: youtu.be/41fctoKS4nU. Moreover, tutorial for general compilation of SDTrimSP and use of its local GUI is accessible at: youtu.be/DwTXVmtTUzw.

Example datasets used in this work are provided within the same GitHub repository in the *example* directory. Data from the subfolder *example/static_N_into_V* can be uploaded directly to the static analysis section of the web application. Data from *example/dynamic_N_into_V* are intended for use in the dynamic analysis section, while data from *example/crystalline_N_into_V* can also be uploaded to the static section and contain results for N implantation into V for different crystallographic surface orientations.

## 8. References

1. Mutzke, A., Toussaint, U. v, Eckstein, W., Dohmen, R. & Schmid, K. SDTrimSP Version 7.00. (2024).
2. Hofsäss, H., Zhang, K. & Mutzke, A. Simulation of ion beam sputtering with SDTrimSP, TRIDYN and SRIM. *Appl Surf Sci* 310, 134–141 (2014).
3. Szabo, P. S. *et al.* Dynamic potential sputtering of lunar analog material by solar wind ions. *Astrophys J* 891, 100 (2020).
4. Brötzner, J. *et al.* Solar wind erosion of lunar regolith is suppressed by surface morphology and regolith properties. *Commun Earth Environ* 6, 560 (2025).
5. Krieger, K. *et al.* Scrape-off layer and divertor physics: Chapter 5 of the special issue: on the path to tokamak burning plasma operation. *Nuclear Fusion* 65, 043001 (2025).
6. Lorentzon, M. *et al.* Growth mechanisms and mechanical response of 3D superstructured cubic and hexagonal Hf1-xAlxN thin films. *Acta Mater* 121680 (2025).
7. Szabo, P. S. *et al.* Graphical user interface for SDTrimSP to simulate sputtering, ion implantation and the dynamic effects of ion irradiation. *Nucl Instrum Methods Phys Res B* 522, 47–53 (2022).
8. Badia, R. M. Frontiers in Scientific Workflows: Pervasive Integration with HPC. (2024).
9. Lebreton, A. *et al.* Control of microstructure and composition of reactively sputtered vanadium nitride thin films based on hysteresis curves and application to microsupercapacitors. *Journal of Vacuum Science & Technology A* 42, (2024).
10. García, J. A. *et al.* Surface mechanical effects of nitrogen ion implantation on vanadium alloys. *Surf Coat Technol* 158, 669–673 (2002).
11. Zheng, W., Sun, Z., Gu, Z., Wu, X. & Niu, L. Recent Advances in Vanadium-Based Cathode Materials for Aqueous Zinc-Ion Batteries: from Fundamentals to Practical Applications. *Adv Mater Technol* 2500320 (2025).
12. Ong, S. P. *et al.* Python Materials Genomics (pymatgen): A robust, open-source python library for materials analysis. *Comput Mater Sci* 68, 314–319 (2013).
13. Ziegler, J. F. & Biersack, J. P. The stopping and range of ions in matter. in *Treatise on heavy-ion science* 93–129 (Springer, 1985).

14. Ziegler, J. F., Ziegler, M. D. & Biersack, J. P. SRIM–The stopping and range of ions in matter (2010). *Nucl Instrum Methods Phys Res B* 268, 1818–1823 (2010).

15. Gražulis, S. *et al.* Crystallography Open Database–an open-access collection of crystal structures. *Applied Crystallography* 42, 726–729 (2009).

16. Pokroy, B. *et al.* Atomic structure of biogenic aragonite. *Chemistry of materials* 19, 3244–3251 (2007).

17. Jain, A. *et al.* The materials project: Accelerating materials design through theory-driven data and tools. in *Handbook of Materials Modeling: Methods: Theory and Modeling* 1751–1784 (Springer, 2020).

18. Nordlund, K., Djurabekova, F. & Hobler, G. Large fraction of crystal directions leads to ion channeling. *Phys Rev B* 94, 214109 (2016).

19. Markelj, S. *et al.* Unveiling the radiation-induced defect production and damage evolution in tungsten using multi-energy Rutherford backscattering spectroscopy in channeling configuration. *Acta Mater* 263, 119499 (2024).